\documentclass[sigconf,authorversion,nonacm]{acmart} %
\AtBeginDocument{%
  }

\copyrightyear{2026}
\acmYear{2026}
\setcopyright{cc}
\setcctype{by}
\acmConference[SIGIR '26]{Proceedings of the 49th International ACM SIGIR Conference on Research and Development in Information Retrieval}{July 20--24, 2026}{Melbourne, VIC, Australia}
\acmBooktitle{Proceedings of the 49th International ACM SIGIR Conference on Research and Development in Information Retrieval (SIGIR '26), July 20--24, 2026, Melbourne, VIC, Australia}
\acmDOI{10.1145/3805712.3809909}
\acmISBN{979-8-4007-2599-9/2026/07}

\usepackage{bm}
\let\vec\bm

\begin{document}

\title{ZoRRO: A Zero-Weight Personalized Recommender System for Scalable News Recommendation}

\author{Johannes Kruse}
\email{jkru@dtu.dk}
\orcid{0009-0007-5830-0611}
\affiliation{%
  \institution{Technical University of Denmark}
  \city{Kongens Lyngby}
  \country{Denmark}
}
\additionalaffiliation{ %
  \institution{JP/Politikens Media Group}
  \city{Copenhagen}
  \country{Denmark}
}

\author{Ryotaro Shimizu}
\email{ryotaro.shimizu@zozo.com
}
\orcid{0000-0002-4841-1824}
\affiliation{%
  \institution{ZOZO Research}
  \city{Tokyo}
  \country{Japan}
}

\author{Kasper Lindskow}
\email{kasper.lindskow@jppol.dk}
\orcid{0009-0004-6412-0930}
\affiliation{%
  \institution{JP/Politikens Media Group}
  \city{Copenhagen}
  \country{Denmark}
}

\author{Jon Tofteskov}
\email{jon.tofteskov@jppol.dk}
\orcid{0009-0003-4855-2904}
\affiliation{%
  \institution{JP/Politikens Media Group}
  \city{Copenhagen}
  \country{Denmark}
}

\author{Michael Riis Andersen}
\email{miri@dtu.dk}
\orcid{0000-0002-7411-5842}
\affiliation{%
  \institution{Technical University of Denmark}
  \city{Kongens Lyngby}
  \country{Denmark}
}

\author{Julian McAuley}
\email{jmcauley@eng.ucsd.edu}
\orcid{0000-0003-0955-7588}
\affiliation{%
  \institution{University of California San Diego}
  \city{La Jolla}
  \state{California}
  \country{USA}
}

\author{Jes Frellsen}
\email{jefr@dtu.dk}
\orcid{0000-0001-9224-1271}
\affiliation{%
  \institution{Technical University of Denmark}
  \city{Kongens Lyngby}
  \country{Denmark}
}
\additionalaffiliation{ %
  \institution{Pioneer Centre for Artificial Intelligence}
  \city{Copenhagen}
  \country{Denmark}
}

\renewcommand{\shortauthors}{Johannes Kruse et al.}

\begin{abstract}
    In this paper, we present ZoRRO (Zero-Weight Personalized Recommender System), a zero-weight and training-free framework for personalized news recommendation designed for scalable real-world deployment. 
    We show that ZoRRO outperforms strong neural baselines in offline ranking evaluations and delivers click-through rate performance in online A/B testing that is nearly on par with a state-of-the-art deep learning model, while operating more than \(600\times\) faster. 
    Our experiments reveal gaps between offline and online performance, and show that models with similar click-through rate (CTR) outcomes can produce markedly different recommendation distributions, influencing the overall news flow.
    These findings position ZoRRO as a practical and efficient solution for large-scale news recommendation and highlight the importance of evaluating recommender systems using metrics beyond accuracy alone.
    Our code is available at \url{https://github.com/johanneskruse/zorro}.
\end{abstract}

\ccsdesc[500]{Information systems~Recommender systems}

\keywords{recommender systems, news recommendation, scalability, online A/B testing}

\maketitle

\section{Introduction}
Unlike other domains, news platforms face unique challenges due to the constant influx of new articles and their short lifespan. This dynamic environment demands recommendation systems that quickly adapt to fresh content and evolving user interests, making the cold-start problem especially severe for both items and users \citep{Das2007, Wu2020MIND, kruse24_ebnerd, kruse24_recsys_challenge, kruse_nails_recsys2025}. 
Despite advances in deep learning-based recommender systems (DLRS), these methods often require substantial computational resources, specialized expertise, and continuous maintenance \citep{Liu2022_hercules, Verachtert2023, Balasubramanian2021, Argyriou2020, Gupta2020}. 
In practice, simpler approaches often prevail, as complex models are harder to implement and sustain, limiting adoption. Moreover, while DLRS offer great flexibility, their benefits and reported gains are not always justified \citep{Dacrema2019, Rendle-2020-ncf-vs-mf, steck2019ease, lian2020lighrec, li2024embedding_compression}. 
Consequently, there is growing interest in practical, scalable approaches that balance effectiveness with operational efficiency \citep{kruse24_ebnerd}. News recommenders must efficiently handle rapid content turnover, address cold-start issues, and deliver timely, personalized recommendations without heavy computational demands \citep{Wu2020MIND, Wu-2023-personalized-NRS}.

To meet these challenges, we propose ZoRRO (Zero-Weight Personalized Recommender System), a simple yet effective framework optimized for the dynamic nature of news recommendation. ZoRRO combines article recency with representation-based similarity and can incorporate article representations generated through different methods, including one-hot category encodings and pre-trained language models \citep{devlin2019bert}. By emphasizing simplicity and efficiency, ZoRRO enables seamless integration of new content and rapid adaptation to shifting user preferences without model retraining. In offline experiments, ZoRRO outperforms strong neural baselines on ranking metrics. In an online A/B test, it delivers CTR close to a state-of-the-art neural model while operating with substantially lower latency and implementation complexity. A beyond-accuracy analysis further shows that systems with similar headline performance can induce meaningfully different recommendation distributions in practice, emphasizing the need to assess recommendation behavior beyond traditional ranking metrics.
Our contributions in this paper are:
\begin{itemize}
    \item We introduce ZoRRO, a zero-weight, training-free recommendation framework for news recommendation that combines recency with representation-based similarity in a form that is simple to implement and efficient to serve.
    \item We show through large-scale offline experiments and systems measurements that a well-tuned lightweight method can outperform or match strong neural baselines while being more than $600\times$ faster at inference.
    \item We validate the approach in a live online A/B test, showing that strong offline ranking performance does not automatically translate into the best online CTR, and thereby highlighting the importance of evaluating practical recommenders beyond offline benchmarks alone.
    \item We provide an offline and online beyond-accuracy analysis showing that models with similar CTR or ranking performance can lead to different topical and sentiment distributions, with direct implications for reliability, editorial alignment, and deployment decisions.
\end{itemize}

\section{Related work}
Most research in recommender systems has concentrated on increasingly complex neural architectures designed to model user behavior and item relationships in great detail \citep{Raza-2022-news-survey, Wu-2023-personalized-NRS}.
Yet, several studies question whether this growing complexity translates into practical benefits. 
For instance, \citet{Dacrema2019} conducted a large-scale reassessment of DLRS and showed that many neural architectures fail to outperform well-tuned shallow methods such as matrix factorization or \(k\)-nearest neighbors. 
In a complementary analysis, \citet{Rendle-2020-ncf-vs-mf} revisited the comparison between Neural Collaborative Filtering (NCF) and matrix factorization, emphasizing that the additional complexity of deep models rarely yields practical benefits. They argue that in production environments, where latency, scalability, and retrieval efficiency are critical, simpler, well-optimized models often remain the preferred choice. 

\citet{steck2019ease} introduced EASE, a closed-form linear autoencoder that achieves competitive top-\(N\) recommendation performance without iterative training, 
while \citet{lian2020lighrec} proposed LightRec, a compact recurrent model that reduces parameter counts substantially without sacrificing accuracy.
These studies underscore the potential of efficient, low-complexity designs for scalable recommendation. 
Building on this line of work, we propose ZoRRO.
Whereas existing lightweight models reduce complexity through compression or architectural simplification, ZoRRO is a lightweight and computationally efficient framework for personalized news recommendations that models user interests with minimal overhead through a zero-weight, training-free design.

\section{Methodology}
\label{sec:methodology}
For a candidate article $c$, ZoRRO computes a relevance score \( r(c) \in \mathbb{R} \) quantifying the predicted user engagement, based on two components: (1) The \textit{intrinsic relevance} (\( \phi \)), which captures an article’s inherent importance, denoted \( \phi_{\operatorname{c}}(c) \in \mathbb{R} \) for a candidate article \( c \) and \( \phi_{\operatorname{h}}(h) \in \mathbb{R} \) for an article \( h \) in the user’s history \( H = [h_1, \dots, h_M] \). The intrinsic relevance may depend on features such as \textit{recency}, \textit{popularity}, or \textit{user-specific preferences}. (2) The \textit{relational relevance} (\( \tau \)), which measures the relationship or similarity between a candidate article \( c \) and a historical article \( h \in H \), denoted \( \tau(c, h) \in \mathbb{R} \).
Given a user’s history \( H \), the relevance score for a candidate article \( c \) is computed as:
\begin{equation}
    r(c) = \phi_{\operatorname{c}}(c) \sum_{h \in H} \tau(c, h) \phi_{\operatorname{h}}(h).
\end{equation}
For a set of candidate articles \( C = [c_1, \dots, c_N] \), the relevance scores can be expressed in vector form as:
\begin{equation}
    \label{eq:zorro_vector_form}
    \vec{r}(C) = \vec{\phi}_{\operatorname{c}}(C) \odot \left( \vec{\tau}(C, H)  \vec{\phi}_{\operatorname{h}}(H) \right),
\end{equation}
where \( \odot \) denotes the Hadamard product and \( \vec{r}(C)_i = r(c_i) \) and \( \vec{\tau}(C, H) \in \mathbb{R}^{N \times M} \). 

\paragraph{Intrinsic relevance}
\label{sec:intrinsic_relevance_method}
We model \textit{intrinsic relevance} using an exponential decay function that assigns higher weight to recent articles:
\begin{equation}
    \label{eq:exponential_decay_function}
    \phi_{\operatorname{c}}(x) = e^{-\lambda_{\operatorname{c}} t(x)} 
    \quad \textrm{and}
    \quad 
    \phi_{\operatorname{h}}(x) = e^{-\lambda_{\operatorname{h}} t(x)},
\end{equation}
where \( \lambda_{\operatorname{c}} \) and \( \lambda_{\operatorname{h}} \) are decay rate parameters controlling how quickly weights decrease over time, and \( t(x) \) denotes the time since article \( x \) was published (in hours). 
This formulation balances simplicity and effectiveness, capturing the importance of timeliness while relying on only one tunable parameter, which allows easy adaptation across time scales and ensures computational efficiency.

\paragraph{Relational relevance}
\label{sec:relational_relevance_method}
To compute the \textit{relational relevance} \( \tau(c, h) \) between a candidate article \( c \) and a historical article \( h \), we measure the distance between their representations using cosine similarity \( \cos(\cdot) \).
Each article is represented in two ways: (1) a document embedding \( \vec{e} \) generated from a language model (e.g., BERT~\citep{devlin2019bert}) that captures textual semantics, and (2) a one-hot encoded vector \( \vec{v} \) indicating its category.
The final relational relevance is defined as the sum of content-based and category-based similarities:
\begin{equation}
    \tau(c,h) = \cos(\vec{e}_c, \vec{e}_h) + \cos(\vec{v}_c, \vec{v}_h).
\end{equation}
We use cosine similarity because it provides a simple and consistent way to compare heterogeneous article representations within the same framework. Importantly, the framework is not tied to this specific choice. Both the similarity function and the article representations can be replaced without changing the overall formulation. For example, the one-hot category representation could be substituted with a learned category embedding, while preserving the same relational relevance framework.

\section{Experiments}
\label{sec:experiments}

\subsection{Experimental setup (offline)}
\label{sec:experimental_setup}

\subsubsection{Dataset}
We use the Ekstra Bladet News Recommendation Dataset (EB-NeRD)~\citep{kruse24_ebnerd}\footnote{\url{https://recsys.eb.dk}} for offline evaluation, containing over 37 million impression logs from 1 million users and 125,000 unique articles. The dataset spans diverse topics, with entertainment (ENT), news (NWS), crime (CRM), and sports (SPT) accounting for \(78\%\) of impressions, alongside smaller categories such as personal finance (PFI), auto-generated content (AGC), and miscellaneous (MSC).

\subsubsection{Evaluation protocol}
We report performance on the EB-NeRD large dataset using its official evaluation framework. For experiments on the hidden test set, we concatenate the training and validation splits, reserving the last 24 hours of data for early stopping and model checkpointing of baselines. The evaluation includes standard ranking metrics, area under the ROC curve (AUC), mean reciprocal rank (MRR), and normalized discounted cumulative gain (nDCG)~\cite{Wu2020MIND, Wu-2023-personalized-NRS}. 
We also consider a preliminary beyond-accuracy framework~\citep{kruse24_recsys_challenge}, adapted to news recommendation, consisting of diversity, serendipity, coverage, and novelty, which capture how dissimilar the recommended items are, how unexpected yet relevant they are, how broadly the recommendation space is utilized across users, and how strongly the system favors less exposed items, respectively~\citep{Smyth2001_intralistdiversity, ge2010_serendipity_coverage, Kaminskas2016}. In addition, we report descriptive analyses of sentiment and category distributions.

\subsubsection{Baselines}
Following prior work~\citep{iana-etal-2024-train, qi2022newsrecommendationcandidateawareuser, li-etal-2022-miner, Wu2020MIND}, we benchmark ZoRRO against widely used baselines: NRMS~\citep{wu2019-nrms}, LSTUR~\citep{an2019-lstur}, and NPA~\citep{wu2019-npa}, which rely on self-attention, GRU-based sequential modeling, and personalized attention, respectively. We also include two heuristic baselines: Popular, ranking articles by recent clicks, and Publish, ranking them by publication time.

\subsubsection{Implementation details}
We tune all models with Optuna~\citep{optuna_2019} using the Tree-Structured Parzen Estimator (TPE)~\citep{watanabe_tpe_tutorial2023} to optimize AUC on the EB-NeRD small training and validation split. Training uses \(5\) epochs, batch size \(32\), and a negative sampling ratio of \(K=4\)~\citep{wu2019-npa}. NRMS uses 16 attention heads (16-dimensional each), while user embeddings are 400-dimensional for LSTUR and NPA. All three models use Adam~\citep{adam_kingma2014} with learning rate \(10^{-4}\), L2 regularization \(10^{-4}\), and \(20\%\) dropout. All experiments were conducted on Amazon EC2 instances (g3.4xlarge with NVIDIA M60 GPU and p3.2xlarge with NVIDIA Tesla V100 GPU).

\subsection{Offline experiments}
\label{sec:offline_experiments}
\subsubsection{Ranking}
\label{sec:ebnerd-large-ranking}
Table~\ref{tab:ranking_metrics_ebnerd} presents the hidden test set results. ZoRRO achieves the best performance across all ranking metrics, with NRMS as the strongest neural baseline. For example, ZoRRO reaches an AUC of 63.80 compared with 62.46 for NRMS, while also improving MRR, nDCG@5, and nDCG@10. While article recency contributes to ZoRRO's effectiveness, the weak performance of Publish (51.68), which ranks purely by publication time, shows that naive recency-based ranking alone is insufficient. These results suggest that strong ranking performance in large-scale news recommendation can be achieved without relying on a complex trained model.

\begin{table}[tbp]
    \centering
    \caption{Performance comparison on the hidden test set.}
    \vspace{-0.5em}
    \resizebox{0.425\textwidth}{!}{%
    \vspace{-0.9em}
    \begin{tabular}{lccccc}
    \toprule
    Method & AUC $\uparrow$ & MRR $\uparrow$ & nDCG@5 $\uparrow$ & nDCG@10 $\uparrow$ \\ \midrule
    ZoRRO       & \textbf{63.80} & \textbf{41.28} & \textbf{46.40} & \textbf{52.58} \\ 
    NRMS        & 62.46 & 40.79 & 45.56 & 52.19 \\ %
    LSTUR       & 59.81 & 38.27 & 42.76 & 49.91 \\ %
    NPA         & 61.57 & 40.46 & 45.05 & 51.79 \\ %
    \midrule
    Popular     & 59.70 & 37.74 & 42.36 & 49.65 \\
    Publish     & 51.68 & 31.61 & 34.84 & 43.19 \\ 
    \bottomrule
    \end{tabular}
    }
    \label{tab:ranking_metrics_ebnerd}
\end{table}

\subsubsection{Beyond-accuracy (offline)}
\label{sec:offline_ba_analysis}
Table~\ref{tab:beyond_accuracy_results_offline} shows that the models exhibit clear trade-offs beyond ranking accuracy. LSTUR achieves the highest diversity (0.74), while NPA attains the highest coverage (0.65), indicating broader use of the candidate space across users. ZoRRO achieves competitive coverage (0.55) and lower novelty than NRMS, suggesting that its recommendations are less exploratory but more concentrated on recent news content. Overall, the results show that models with comparable ranking quality can induce notably different recommendation behaviors~\citep{kruse23_next_gen_recsys, kruse2025design_choices}. 
Because this offline evaluation reflects a static snapshot of the news cycle, we extend the analysis to the online setting in Section~\ref{sec:online_ba_analysis}.

\begin{table*}[tbp]
    \centering
    \caption{
        Offline beyond-accuracy results on the hidden test set. We report diversity, serendipity, coverage, novelty, sentiment distribution, and category distribution for the top-5 recommendations. Results are computed using a beyond-accuracy candidate set consisting of the 250 most recent articles published during the test period.
    }
    \label{tab:beyond_accuracy_results_offline}
    \vspace{-0.9em}
    \begin{tabular}{lcccc rrr rrrrrrr}
        \toprule
        & \multicolumn{4}{c}{Beyond-accuracy} & \multicolumn{3}{c}{Sentiment (\%)} & \multicolumn{7}{c}{Category (\%)} \\
        \cmidrule(lr){2-5} \cmidrule(lr){6-8} \cmidrule(lr){9-15}
        Method & Diversity & Serendipity & Coverage & Novelty & NEG & NEU & POS & CRM & NWS & SPT & ENT & PFI & AGC & MSC \\
        \midrule
        ZoRRO   & 0.58  & 0.70  & 0.55 & 5.1  & 69.4  & 23.4  & 7.2   & 18.4  & 46.9  & 22.1  & 10.8  & 1.0   & 0.0  & 0.8  \\ 
        NRMS    & 0.68  & 0.77  & 0.53 & 8.1  & 45.7  & 18.2  & 36.1  & 12.5  & 18.7  & 29.2  & 9.5   & 4.3   & 18.3 & 7.5  \\ 
        LSTUR   & 0.74  & 0.79  & 0.56 & 5.8  & 58.8  & 6.8   & 34.4  & 30.1  & 24.6  & 32.1  & 6.8   & 0.2   & 1.1  & 5.1  \\ 
        NPA     & 0.65  & 0.75  & 0.65 & 6.7  & 62.8  & 10.3  & 26.9  & 13.6  & 22.0  & 26.4  & 1.2   & 1.6   & 10.4 & 24.8 \\ 
        Popular & 0.84  & 0.79  & 0.02 & 3.1  & 40.0  & 40.0  & 20.0  & 20.0  & 20.0  & 40.0  & 0.0   & 20.0  & 0.0  & 0.0   \\ 
        Random  & 0.76  & 0.81  & 1.00 & 11.1 & 39.6  & 29.2  & 31.2  & 9.6   & 16.0  & 16.8  & 8.4   & 3.6   & 38.0 & 7.2  \\ 
        \bottomrule
    \end{tabular}
\end{table*}

\subsubsection{Inference speed and parameter efficiency}
\label{sec:inference_memory_analysis}
As shown in Table~\ref{tab:throughput_comparison}, ZoRRO achieves more than \(600\times\) higher throughput than NRMS, LSTUR, and NPA. With a fixed history length of \(20\) and embedding dimension of \(768\), it processes nearly \(2{,}000\) requests per second with an average latency of \(0.5\) ms per request. In contrast, NRMS, LSTUR, and NPA require more than \(300\) ms per request and substantially larger parameter counts. By relying on precomputed article representations, ZoRRO reduces online computation and keeps the parameter footprint minimal, highlighting its practical advantages in real-time, latency-sensitive settings.

\begin{table}[tbp]
    \centering
    \caption{
    Throughput (requests/s), latency (ms/request), and parameter count for ZoRRO, NRMS, LSTUR, and NPA using 768-dimensional embeddings and a history length of 20. Results are reported as mean $\pm$ std over 1000 calls, each consisting of 100 inference requests.
    }
    \vspace{-0.9em}
    \begin{tabular}{lllr}
    \toprule
    \multicolumn{1}{c}{Method} & \multicolumn{1}{l}{Throughput $\uparrow$} & \multicolumn{1}{l}{Latency $\downarrow$} & \multicolumn{1}{r}{Parameters} \\ \midrule
    ZoRRO       & $\mathbf{1953.91\pm5.53}$  & $\mathbf{0.51\pm0.001}$    & $2$ \\ 
    NRMS        & $2.93\pm0.04$     & $341.44\pm4.90$   & $323{,}152$ \\
    LSTUR       & $2.25\pm0.02$     & $445.14\pm4.69$   & $21{,}047{,}104$ \\ 
    NPA         & $2.62\pm0.16$     & $384.61\pm39.81$  & $20{,}405{,}304$ \\
    \bottomrule
    \end{tabular}
    \label{tab:throughput_comparison}
\end{table}

\subsubsection{Article embeddings}
In Table~\ref{tab:article_embeddings} we compare several article embedding types, including two embeddings trained on proprietary news data, word2vec~\citep{Mikolov2013} and a contrastive-learning embedding~\citep{gao2022simcse}, as well as two open-source transformer-based models, BERT~\citep{devlin2019bert} and RoBERTa~\citep{liu2019roberta}.
The contrastive-learning embedding performs best for ZoRRO, while word2vec yields the strongest results for NRMS, LSTUR, and NPA, suggesting that different recommendation architectures benefit from different embedding properties. Overall, the results indicate that in-domain embeddings provide a stronger representation of news articles than the more general-purpose transformer embeddings considered here.

\begin{table}[tbp]
    \centering
    \caption{
    Performance with different article embeddings: (1) word2vec~\citep{Mikolov2013}, (2) contrastive learning~\citep{gao2022simcse}, (3) BERT~\citep{devlin2019bert}, and (4) RoBERTa~\citep{liu2019roberta}. Models are evaluated by AUC. Results are reported as mean $\pm$ std over three runs.
    }
    \vspace{-0.9em}
    \begin{tabular}{lcccc}
        \toprule
        AUC $\uparrow$ & ZoRRO & NRMS & LSTUR & NPA \\
        \midrule
        word2vec      & $62.1\pm0.0$        & $\mathbf{61.4\pm0.2}$ & $\mathbf{63.0\pm0.4}$ & $\mathbf{62.9\pm0.2}$ \\
        contrastive   & $\mathbf{63.3\pm0.0}$ & $59.9\pm0.4$           & $62.0\pm0.4$           & $62.2\pm0.1$ \\
        BERT          & $60.8\pm0.0$        & $59.5\pm0.3$           & $59.5\pm0.7$           & $60.5\pm0.8$ \\
        RoBERTa       & $60.2\pm0.0$        & $61.0\pm0.4$           & $60.7\pm0.5$           & $62.4\pm0.2$ \\
        \bottomrule
    \end{tabular}
    \label{tab:article_embeddings}
\end{table}

\subsubsection{Influence of history size}
\begin{figure}
    \centering
    \includegraphics[width=0.85\linewidth]{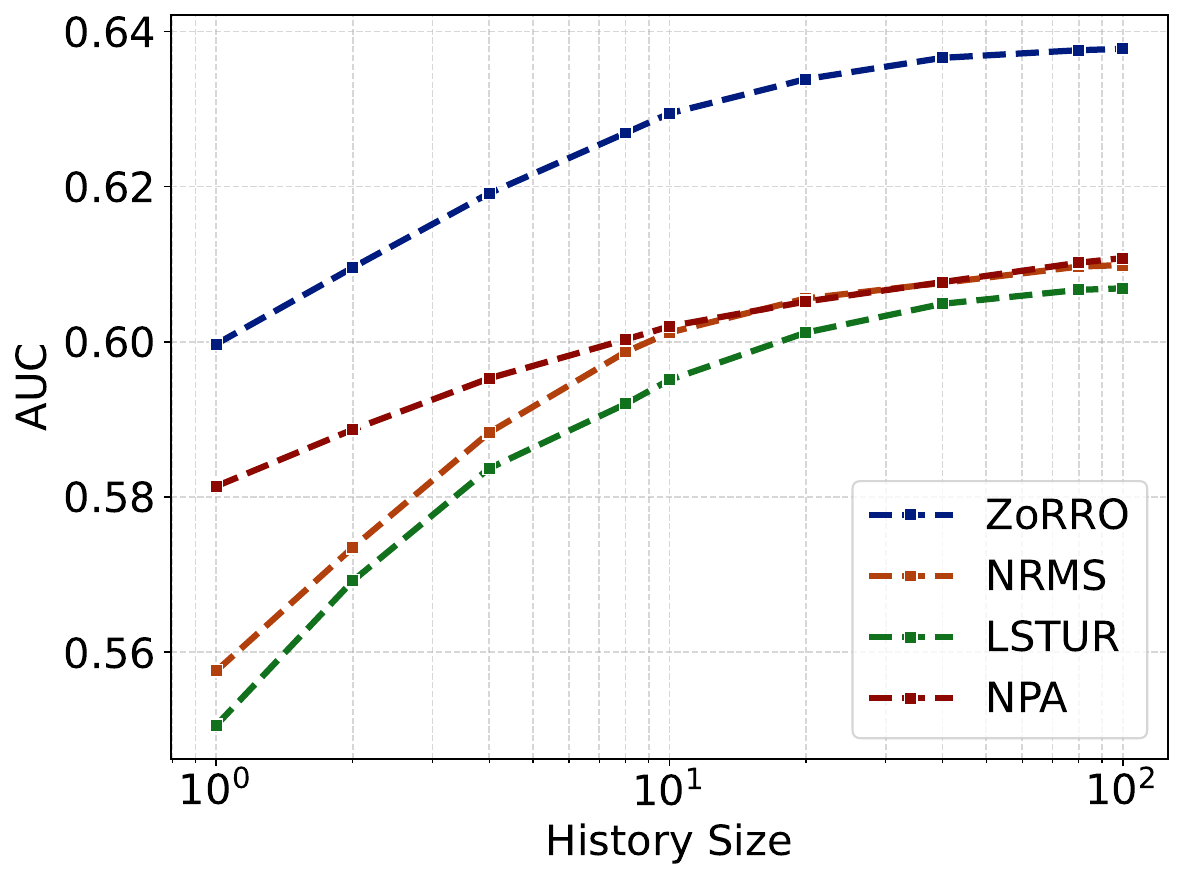}
    \vspace{-0.9em}
    \caption{Effect of history size on AUC performance.}
    \label{fig:history_size_auc_correlation}
    \Description[AUC as function of history size]{Effect of history size on ZoRRO's AUC performance.}
\end{figure}

We varied user click history length from \(1\) to \(100\) interactions to examine its effect on performance (Figure~\ref{fig:history_size_auc_correlation}). LSTUR and NPA were evaluated with masked user IDs to avoid bias from user-specific identifiers. AUC increases with history length but eventually plateaus, indicating diminishing returns and highlighting the trade-off between capturing richer histories and maintaining data coverage for cold-start users.

\subsubsection{Intrinsic and relational relevance}
\label{sec:intrinsic_relational_relevance_exp}
\begin{figure}
    \centering
    \includegraphics[width=0.85\linewidth]{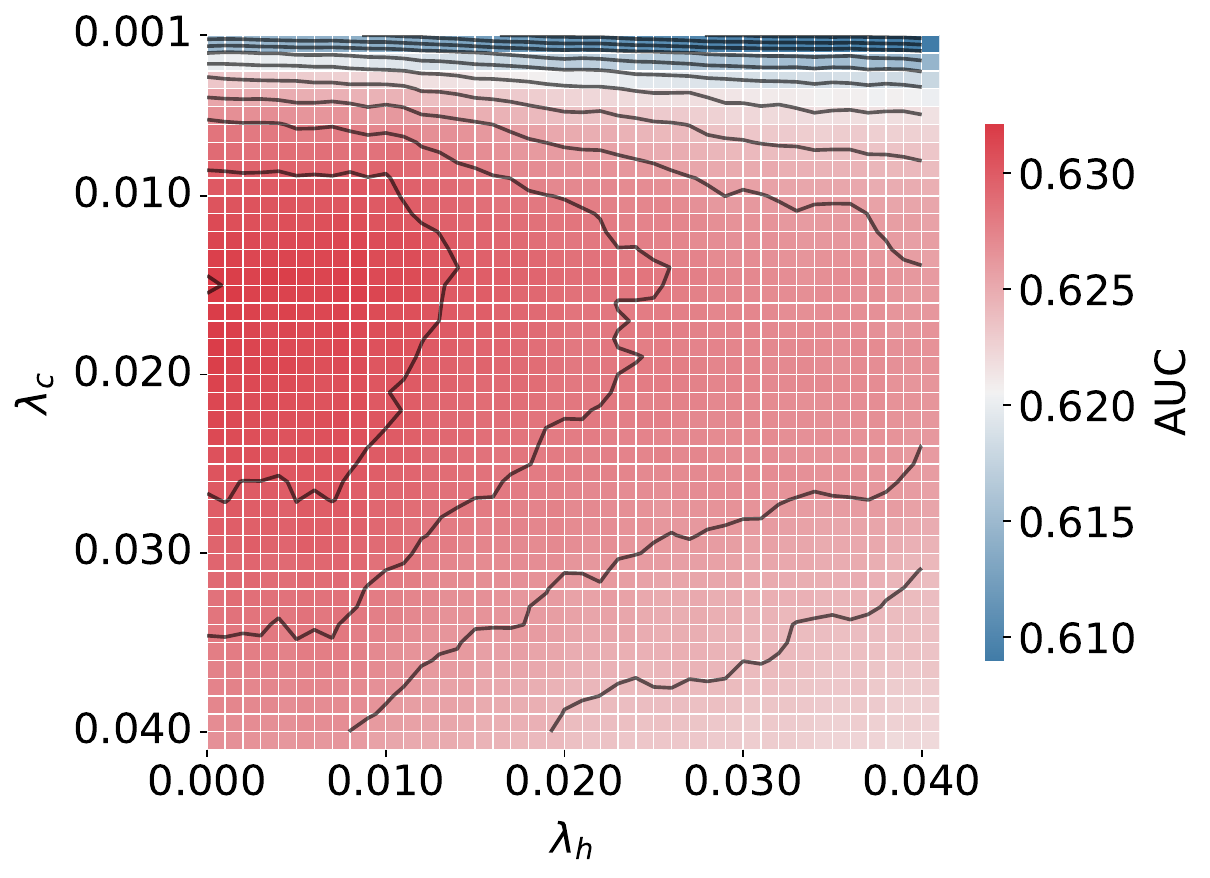}
    \vspace{-0.9em}
    \caption{ZoRRO's AUC as a function of $\lambda_h$ and $\lambda_c$.}
    \label{fig:sensitivity_analysis_heatmap}
    \Description[Heatmap]{Sensitivity analysis using heatmap showing AUC performance as function of $\mathbf{\lambda_h}$ and $\mathbf{\lambda_c}$.}
\end{figure}

We analyze the effect of ZoRRO’s \textit{intrinsic} and \textit{relational relevance} components. For intrinsic relevance, modeled using an exponential decay function, a sensitivity analysis over \(\lambda_c\) and \(\lambda_h\) (Figure~\ref{fig:sensitivity_analysis_heatmap}) identifies the best setting at \(\lambda_c = 0.015\) and \(\lambda_h = 0.0\). This suggests that recency on the candidate side is beneficial, whereas temporal decay over the user history does not improve AUC in this setting.
For relational relevance, Table~\ref{tab:relational_relevance} shows that both semantic and categorical representations contribute to performance. Using both the document embedding \(\vec{e}\) and the category vector \(\vec{v}\) yields the best overall results. Removing the categorical component slightly reduces AUC from \(63.30\) to \(62.39\), while relying on the categorical component alone lowers AUC further to \(58.91\). This indicates that semantic similarity is the stronger signal, but that categorical information provides a complementary improvement.

\begin{table}[tbp]
    \centering
    \caption{ZoRRO performance under varying relational relevance features.}
    \vspace{-0.9em}
    \resizebox{0.47\textwidth}{!}{%
    \begin{tabular}{c c c c c c }
         \toprule
         $\vec{e}$ & $\vec{v}$  & AUC $\uparrow$ & MRR $\uparrow$ & nDCG@5 $\uparrow$ & nDCG@10 $\uparrow$ \\ \midrule
         \checkmark & \checkmark    & \textbf{63.30} & \textbf{40.49} & \textbf{45.41} & \textbf{51.74} \\ 
         \checkmark &               & 62.39 & 40.45 & 45.08 & 51.53 \\
                    & \checkmark    & 58.91 & 36.52 & 41.09 & 48.17 \\
        \bottomrule
    \end{tabular}%
    }
    \label{tab:relational_relevance}
\end{table}

\begin{table*}[tbp]
    \centering
    \caption{
        Online beyond-accuracy results during the A/B test period. We report diversity, serendipity, coverage, novelty, sentiment distribution, and category distribution for the top-5 recommendations. Results are computed using aggregated candidate lists consisting of the 250 most popular recently published articles during the test period.
    }
    \label{tab:beyond_accuracy_results_online}
    \vspace{-0.9em}
    \begin{tabular}{lcccc rrr rrrrrrr}
        \toprule
        & \multicolumn{4}{c}{Beyond-accuracy} & \multicolumn{3}{c}{Sentiment (\%)} & \multicolumn{7}{c}{Category (\%)} \\ 
        \cmidrule(lr){2-5} \cmidrule(lr){6-8} \cmidrule(lr){9-15}
        Method  & Diversity & Serendipity & Coverage & Novelty & NEG & NEU & POS & CRM & NWS & SPT & ENT & PFI & AGC & MSC \\
        \midrule
        ZoRRO   & 0.54 & 0.70 & 0.49 & 11.0 & 54.5 & 30.6 & 14.9 & 1.6 & 47.2 & 27.1 & 23.5 & 0.5 & 0.0 & 0.1 \\ 
        NRMS    & 0.72 & 0.77 & 0.55 & 10.6 & 52.2 & 25.9 & 21.9 & 8.2 & 21.3 & 41.1 & 19.3 & 5.7 & 0.0 & 4.4 \\ 
        Popular & 0.79 & 0.78 & 0.23 & 8.9  & 68.6 & 30.0 & 1.4  & 6.1 & 47.7 & 12.0 & 22.1 & 0.3 & 0.0 & 11.8 \\ 
        Random$^\diamond$ & 0.77 & 0.74 & 1.00 & 11.8 & 49.7 & 29.4 & 20.9 & 7.2 & 23.7 & 44.2 & 15.7 & 5.6 & 0.0 & 3.6 \\ 
        \bottomrule
        \multicolumn{15}{l}{\footnotesize $^\diamond$ Random is computed offline by uniformly sampling from the full candidate set and is included for reference.}
    \end{tabular}
\end{table*}

\subsection{Online A/B test}
We conducted a six-day online A/B test from November 12 to 18, 2024, on Ekstra Bladet's website\footnote{\url{https://ekstrabladet.dk}}, comparing ZoRRO, NRMS, and a Most Popular baseline.
All models were served through a FastAPI application on AWS Fargate connected to a PostgreSQL database that continuously updated article content and embeddings. 
The system was deployed across two identical service instances, and users were assigned consistently across variants: \(45\%\) ZoRRO, \(45\%\) NRMS, and \(10\%\) Popular. All methods used a shared rule-based candidate set of the top \(250\) popular recent articles, updated throughout the test period, resulting in \(108\) unique candidate-lists, with articles potentially appearing in multiple updates.

\subsubsection{Click-through rate}
\label{sec:ab_ctr_exp}
As shown in Table~\ref{tab:ctr_results}, NRMS achieved the highest observed CTR \((4.33\%)\), followed closely by ZoRRO \((4.19\%)\), while both outperformed the Popular baseline \((2.96\%)\). Although ZoRRO performs best in the offline ranking evaluation (Table~\ref{tab:ranking_metrics_ebnerd}), this advantage does not fully transfer to the online setting, indicating a mismatch between offline ranking quality and online user response.

\begin{table}[tbp]
    \centering
    \caption{Estimated CTR and CTR variance ($\sigma^2$) for the three recommendation methods: NRMS, ZoRRO, and Popular.}
    \vspace{-0.9em}
    \begin{tabular}{l c c r r}
         \toprule
         Method & CTR $\uparrow$ & $\sigma^2 \downarrow$ & \#Users & \#Impressions
         \\ 
         \midrule
         ZoRRO     & $4.19\%$ & $\mathbf{7.7 \cdot 10^8}$ & $207{,}112$& $1{,}577{,}698$ \\
         NRMS      & $\mathbf{4.33\%}$ & $9.1 \cdot 10^8$ & $208{,}998$& $1{,}586{,}784$ \\
         Popular   & $2.96\%$ & $2.2 \cdot 10^7$ & $45{,}640$  & $337{,}355$ \\
         \bottomrule
    \end{tabular}
    \label{tab:ctr_results}
\end{table}

\subsubsection{Beyond-accuracy (online)}
\label{sec:online_ba_analysis}
Table~\ref{tab:beyond_accuracy_results_online} shows that the models exhibit clear behavioral differences beyond CTR. NRMS achieves higher diversity \((0.72)\) and coverage \((0.55)\) than ZoRRO \((0.54\) and \(0.49)\), indicating that it exposes users to a broader set of items. ZoRRO, in contrast, produces a more concentrated recommendation profile. This is particularly visible in the category distribution, where ZoRRO allocates \(47.2\%\) of recommendations to news, compared with \(21.3\%\) for NRMS, while NRMS places a larger share on sports \((41.1\%)\). Sentiment distributions also differ, with ZoRRO recommending fewer positive articles \((14.9\%)\) than NRMS \((21.9\%)\). Overall, these results show that models with similar CTR can still induce meaningfully different recommendation behaviors in terms of diversity, topical concentration, and sentiment balance.

\subsubsection{Discussion: offline versus online behavior}
\label{sec:offline_online_discussion}
The results highlight a clear gap between offline and online evaluation. Although the EB-NeRD benchmark and the live A/B test share the same publisher context, they differ in important ways: ZoRRO outperforms NRMS on the offline hidden test set (Table~\ref{tab:ranking_metrics_ebnerd}), yet NRMS achieves the highest estimated CTR online (Table~\ref{tab:ctr_results}). Several differences between the two settings may contribute to this mismatch.
Offline evaluation reflects a static, front-page-skewed snapshot of the news cycle~\citep{kruse24_ebnerd}, whereas the deployed system operates on continuously updated candidates further down the page. This may help explain why stronger recency-based behavior is more effective offline than online. Consistent with this, ZoRRO produces temporally narrower recommendation lists than NRMS. The average time difference between the newest and oldest article in the top-5 list is 9h 50m for ZoRRO, compared with 20h 40m for NRMS and 1d 4h for Popular, while a random sample from the candidate set yields an average span of 1d 10h.
The candidate distributions also differ across settings (Tables~\ref{tab:beyond_accuracy_results_offline} and~\ref{tab:beyond_accuracy_results_online}). ZoRRO's online recommendation profile is more concentrated, with lower diversity (0.54 vs.\ 0.72), lower coverage (0.49 vs.\ 0.55), and a stronger focus on news content (47.2\% vs.\ 21.3\%). NRMS, in contrast, produces a broader recommendation distribution that more closely resembles the overall category distribution.
These differences matter in news recommendation because recommender systems shape exposure to topics, sentiments, and editorial and normative objectives~\citep{normalize_2023_first_workshop, normalize_2023_ceur, Einarsson_2025_user_exp}.
Finally, offline AUC and online CTR capture different evaluation targets, and improvements in offline ranking do not necessarily translate directly into higher click-through in a live setting.

\section{Conclusion}
We presented ZoRRO, a zero-weight and training-free recommendation framework for news recommendation that combines recency with representation-based similarity in a form that is simple to implement, efficient to serve, and practical to deploy. In large-scale offline experiments, ZoRRO outperformed or matched strong neural baselines on standard ranking metrics, while achieving more than \(600\times\) higher inference throughput.
We further validated ZoRRO in a live online A/B test, where it achieved competitive CTR but did not outperform NRMS. Together with our offline and online beyond-accuracy analyses, this shows that strong offline ranking performance does not necessarily translate directly into superior online behavior, and that models with similar accuracy can still produce meaningfully different recommendation distributions. Overall, ZoRRO provides a strong practical baseline for real-time news recommendation and highlights the importance of evaluating recommender systems beyond offline accuracy metrics alone.

\begin{acks}
This work was supported by Innovation Fund Denmark (1044-00058B) and the Platform Intelligence in News Project (0175-00014B).
\end{acks}

\balance
\bibliographystyle{ACM-Reference-Format}
\bibliography{main_ref}

\end{document}